\documentclass{aa}
\usepackage{graphicx}
\usepackage{natbib}
\usepackage{afterpage}
\bibpunct{(}{)}{;}{a}{}{,}

\newcommand\changed{}
\newcommand\changedd{}

\title{The frequency-dependence of drifting subpulse patterns}
\author{R. T. Edwards\inst{1} \and B. W. Stappers\inst{2,1}}
\date{}
\institute{Astronomical Institute ``Anton Pannekoek'', 
        University of Amsterdam,
        Kruislaan 403, 1098 SJ Amsterdam, The Netherlands 
  \and
   Stichting ASTRON, Postbus 2, 7990 AA Dwingeloo, The Netherlands}
\offprints{R.~T. Edwards, \email{redwards@astro.uva.nl}}

\abstract{Drifting subpulse patterns in pulsar signals are frequently
interpreted in terms of a model in which a rotating ring of sparks on
the polar cap gives rise to emission from regions of the magnetsophere
connected to the sparks by dipolar magnetic field lines. The spacing
of drift-bands in time depends on the circulation rate of the polar
cap pattern, but to first order the longitudinal phase dependence of
the subpulse modulation should obey a frequency-independent relation
determined by the geometrical configuration in a similar manner to the
polarization position angle. We present here observations at
272--1380~MHz of PSR B0320+39 and PSR B0809+74, both of which show
nearly linear drift in two longitude regions, separated by a region of
reduced modulation and accompanied by a large step in the phase of the
subpulse pattern. We show that the observation of \citet{bkk+81} that
the subpulse spacing for PSR B0809+74 was 1.8 times greater at 102.5
MHz than at 1720 MHz is most likely an artifact of the phase step,
which is only present at high frequencies. The phase steps can be
understood as a consequence of observing overlapping offset images of
the polar cap spark pattern. We also detected more complicated,
frequency-dependent behaviour that would require that the images do
not simply differ by rotation about their centers.  Detailed modelling
of non-axisymmetric refraction or distorted magnetic fields is
suggested as a means of pursuing an explanation for this phenomenon.
}

\begin{document}

\maketitle
\section{Introduction}
It is generally accepted that the radio emission from pulsars is
beamed from its points of origin along local tangents to the dipolar
magnetic field lines, and that the general broadening of pulsar
average profiles with decreasing observing frequency can be understood
as a consequence of the divergence of the field lines and the scaling
of emission frequency with radial distance from the polar cap
(e.g. \citealt{kom70,cor78}). So fundamental is this so-called
radius-to-frequency mapping (RFM) to the usual ways of thinking about
pulsar emission, that many authors appear to assume that it will apply
to longitudinal subpulse spacing ($P_2$) in the same way that it
applies to component separations and average profile width
(e.g. \citealt{rs75,wbs81,bkk+81,bar81,ikl+93}). As noted by
\citet{gk96b}, this is only the case when the line of sight passes
directly over the magnetic pole, and the weaker dependence on
frequency seen in $P_2$ versus the average profile width reported by
\citet{ikl+93} {\changedd and \citet{gggk02}} is just as expected
under the usual model of drifting subpulse patterns as manifestations
of a grazing pass along the edge of a beam consisting of a ring
(``carousel'') of beamlets circulating about the magnetic axis
(e.g. \citealt{rs75}).

\begin{figure}[htb]
\resizebox{0.3\hsize}{!}{\includegraphics{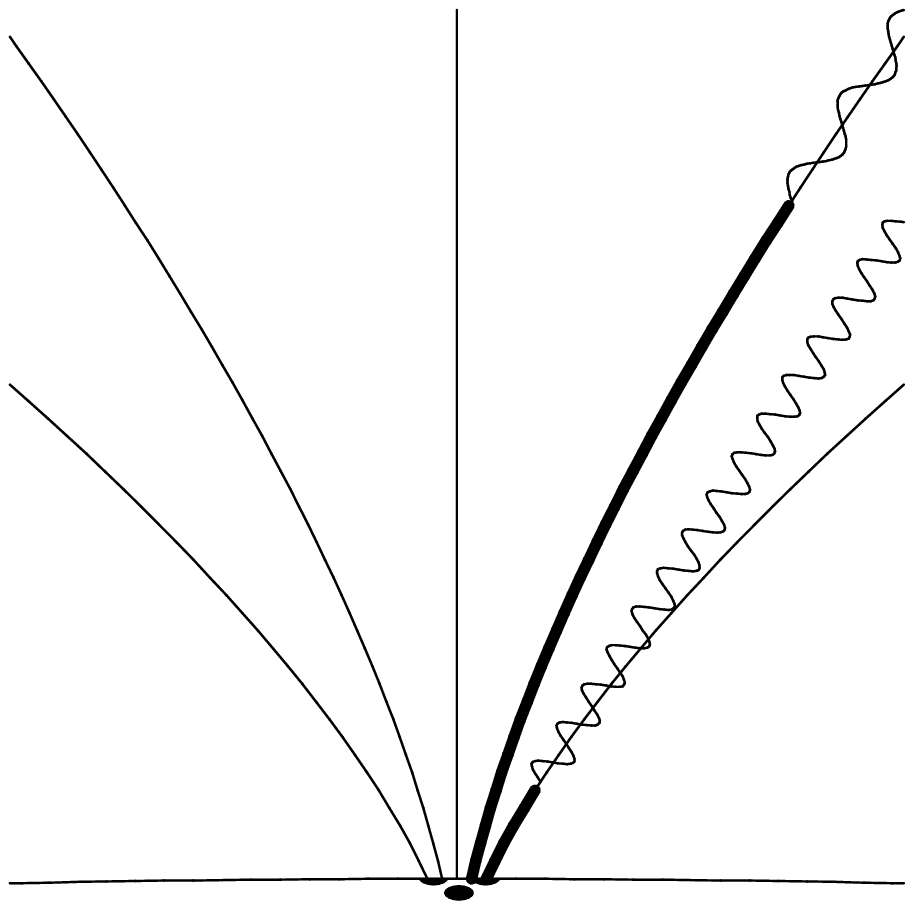}} \\
\vspace{-1cm}
\flushright\resizebox{0.8\hsize}{!}{\includegraphics{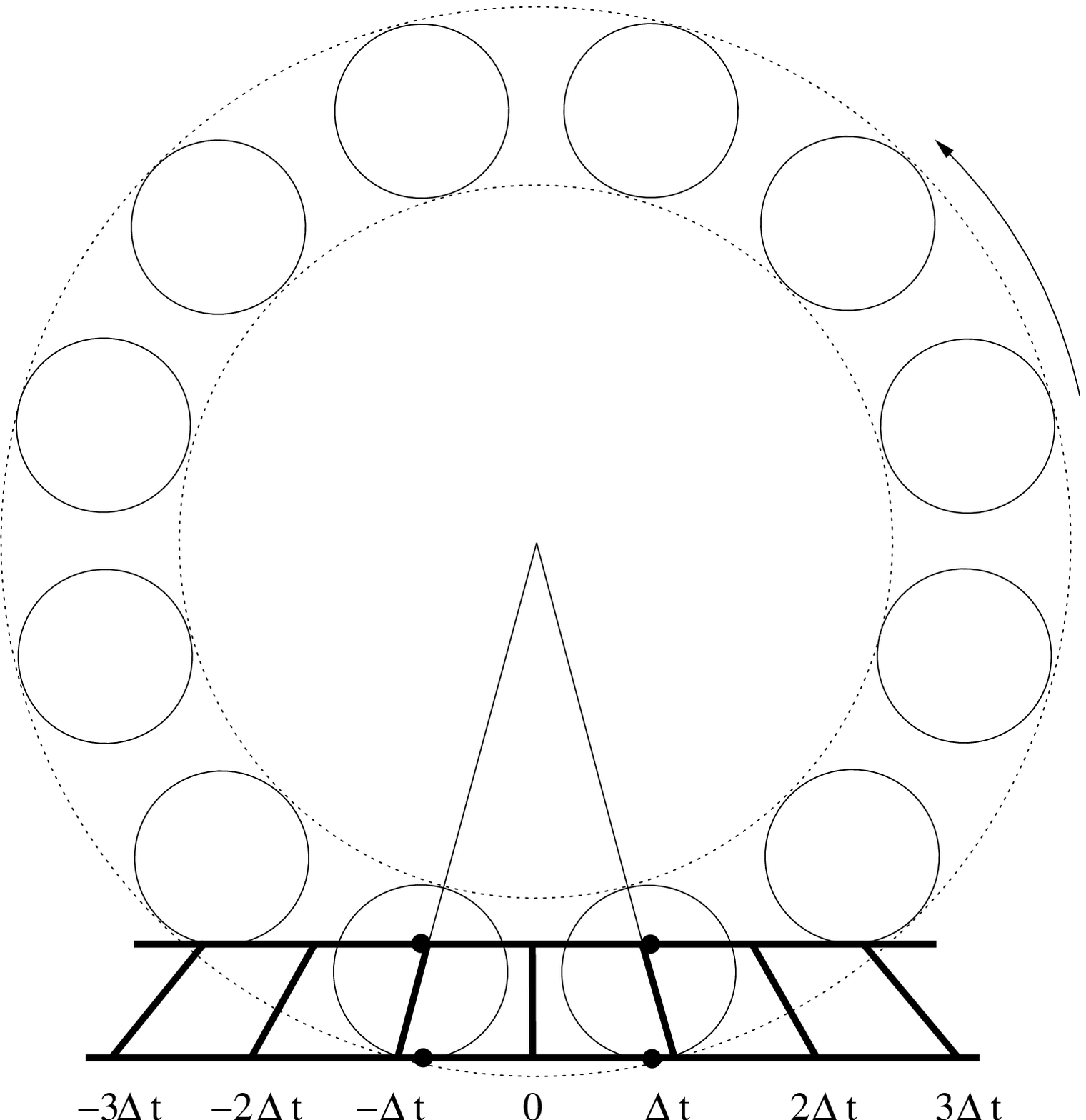}}
\caption{(Top left) Schematic diagram depicting the origin of
simultaneously observable rays (wavy lines) emitted at different
altitudes. The thick lines show the path of plasma flows from the
polar cap excitation pattern, along field lines to the points of
emission of these rays. Due to RFM and the divergence of the field
lines, low frequency observers sample a more poleward path on the
polar cap.  The thin lines show other co-planar field lines, and the
stellar surface. (Bottom right) Diagram of a carousel system on the
polar cap. The solid circles represent a system of sparks circulating
about the magnetic pole. The thick horizontal lines depict the path
sampled by observers receiving radiation emitted from two different
heights in the magnetosphere (the upper line corresponding to a
greater altitude) along a circular path effected by the rotation of
the star under a specific viewing geometry. These lines are connected
at several points to indicate simultaneously observed locations
(neglecting propagation delays), spaced at equal time/longitude
intervals $\Delta t$. The lines emanating from the center of the
carousel indicate two of the twelve lines of zero phase in the
azimuthal modulation, which are seen at the same pulse longitudes
($\pm \Delta t$, chosen for convenience) for both observers. The small
filled circles show the longitudes at which subpulse peaks are
recorded by each observer (for this carousel orientation). Unlike
fiducial phase points, the longitudinal separation between subpulse
peaks increases with the altitude of emission.  The overall pulse
width, as indicated by the intersections of the outer dotted circle
with the thick lines, evolves even more strongly.}
\label{fig:cap}
\end{figure}

\citet{es02} (ES) showed that if one considers the subpulse pattern as
a periodic modulation within any given longitude interval, the {\em
phase} of the modulation as a function of longitude should be {\em
independent} of frequency, given a {\changed
circular
pattern of simply shaped
sparks} (i.e. extended in either magnetic azimuth or latitude but not
in both), and neglecting the effects of rotation.  This is a
consequence of the fact that the transformation between the polar cap
pattern and the beam pattern as radiated from a certain height is a
simple scaling of the polar opening angle (see Fig. \ref{fig:cap}, top
left). The sampling of the polar cap effected at a particular
longitude value at different radio frequencies therefore differs in
magnetic latitude but not in magnetic azimuth, and for a simple
circular carousel this means that the amplitude of the subpulses may
differ but their phase does not.  Figure \ref{fig:cap} (bottom right)
shows the difference between points of zero subpulse phase and points
of peak subpulse intensity (between which $P_2$ is measured) for a
pair of sight-lines on a sample polar cap configuration.  It shows
that the variation with frequency of $P_2$ is entirely a consequence
of the different amplitude windowing caused by the expansion or
contraction of the beam. In contrast, the subpulse phase envelope is
unaffected by the different amplitude windowing and its longitude
dependence should be describable by a simple geometric formula,
independent of frequency (ES).  The possibility therefore exists for a
strong test of the carousel model by examining the longitude- and
frequency-dependence of subpulse phase in much the same way as the now
well-accepted geometric polarization model of \citet{rc69a} was
tested.

Application of phase-based analysis to PSR B0320+39 has shown that at
328~MHz, subpulse drifting occurs in two distinct longitude intervals,
between which there is a phase offset  of
nearly 180\degr\ (\citealt{esv03}; ESvL). The
transition between the regions occurs too suddenly to be fit by the
geometric model, and ESvL suggested the phase step reflected a shift
in dominance between two superposed, phase-offset {\changed images of
an underlying rotating subbeam system. If this is the case, in the
absence of further complicating factors, the underlying phase
envelopes of the two images should exhibit the predicted frequency
independence, and any variations in the measured envelope must be due
to frequency dependence of the relative amplitudes of the images at
any given longitude. If the offset is close to 180\degr at all
longitudes, the phase must always be close to that of the dominant
component (at that longitude) and after accounting for phase steps,
the prediction of frequency independence can be tested.}

Similar behaviour, involving a subpulse phase jump accompanied by
reduced modulation in the vicinity of the jump, has been reported at
1720 and 1420 MHz for PSR B0809+74 \citep{wbs81,pw86}, {\changed
although the quality of these results is insufficient to strongly
constrain the sharpness of the jump, or to probe for other
non-linearities and their variation with frequency. Without this
information it is difficult to distinguish between the multiple
imaging scenario and other possibilites
(e.g. \citealt{bar81,dls+84}).} To investigate this phenomenon further
we have studied PSR B0320+39 and PSR B0809+74 at frequencies between
270 and 1380 MHz.

\section{Observations and Analysis}
We observed PSRs B0320+39 and B0809+74 using the Westerbork Synthesis
Radio Telescope and its pulsar back end, PuMa. The signals from
linearly polarized receptors from 14 25-m dishes were decimated into
bands of 10~MHz width, added (per-band, per-polarization)
according to previously determined phase and amplitude factors, and
fed in to PuMa operating in Mode 1, which acts as a digital
filterbank; for details see \citet{vkv02}. In all cases we used a
configuration such that both the dispersion smearing within each
frequency channel and the output sampling interval were less than or
equal to $409.6$~$\mu$s.  In offline processing total intensity
samples from all frequency channels were combined according to the
\clearpage 
appropriate compensation for interstellar dispersion, and where
necessary  adjacent samples were added to improve the
signal-to-noise ratio. 
Based on the topocentric period predicted using
the TEMPO\footnote{http://pulsar.princeton.edu/tempo/} software
package and published ephemerides, the resultant time-series was
re-arranged into a two-dimensional array of pulse longitude and pulse
number for further analysis. We made observations with bands centered
at 272--302, 328, 382 and 1345--1415 MHz, however for both pulsars,
the results were consistent within their respective uncertainties for
all frequencies between 272 and 382 MHz, so in addition to the 1380~MHz
data, we will present results only from the most sensitive band, which
was 328 MHz in both cases.

For pulsar signals that are dominated by a regular drifting subpulse
pattern, the method of ES offers the best possible
signal-to-noise ratio for the inference of the time- and
longitude-dependence of subpulse phase. However, we find that for both
pulsars, at 1380~MHz the regular drifting subpulses comprise a
relatively minor portion of the total emission, and the time/longitude
decomposition fails for this reason.  To overcome this, in this work
we used an alternative algorithm to ascertain the longitude-dependence
of subpulse amplitude and phase (hereafter, the complex
``envelope''). Specifically, we divided the pulse series in to
$256$-pulse blocks and computed Discrete Fourier Transforms 
along constant-longitude columns of each block to form a series of
complex Longitude-Resolved Fluctuation Spectra (LRFS). If the subpulse
pattern is at least strong and coherent enough to dominate over other
emission in its peak spectral frequency bin, that row of any given
spectrum should differ from that of any other spectrum by a simple
complex factor. The phases of these factors were determined by
cross-correlation with a template envelope, a new template was formed
by adding the peak rows using the measured phases, and the process
repeated until convergence. This is essentially a coarse-grained
version of the algorithm of ES, and has been shown to
produce consistent results in practice (ESvL) for the phase
envelope. 

The subpulse amplitude envelope derived with this method is typically
underestimated since the single Fourier coefficient is not matched to
the true response of the signal. To overcome this, we measured the LRF
power spectrum as the difference between on-pulse power spectra and
off-pulse (noise) spectra, and estimated the variance of the
subpulse-modulated part of the signal as the sum of power spectral
values in a narrow range of frequencies around the peak response
(Parseval's Theorem). Assuming the subpulse signal has a similar
amplitude distribution to that of a sinusoid, we then took the value
$(2v)^{1/2}$ where $v$ is the variance as an estimate of the typical
height of subpulse peaks above the mean in the given longitude bin. As
expected, for low-frequency observations (where the subpulse signal
dominates) the full time/longitude decomposition method (ES)
gave a result consistent with this method.

\section{Results and Discussion}

\subsection{PSR B0320+39}
The average profiles and subpulse amplitude and phase envelopes for
PSR B0320+39 at 328~MHz and 1380~MHz are shown in Fig. \ref{fig:res}.
We used the same observation as ESvL at 328~MHz, and the results are
consistent within the noise (despite using different techniques).  A
sharp jump in subpulse phase is accompanied by an almost complete
attenuation of subpulse amplitude, consistent with destructive
interference between two offset drift patterns with differing
longitudinal amplitude dependences. The presence of an additional
component was the motivating factor for examining the subpulse
properties at high frequency, with the prediction that it should be
preceded by another jump to align it with the extrapolation of the
subpulse phase envelope in the leading part of the profile. The extra
component is clearly detected here, as is its subpulse modulation (Fig
\ref{fig:res} left, longitude $\sim 26$\degr). However, the phase
envelope differs from the prediction, and tends towards almost
stationary modulation in the trailing component. Moreover, the sharp
jump seen at 328~MHz (longitude $\sim 19$\degr) is apparently reduced
to a smoother transition of the opposite sense (seen most clearly in
the middle panel). The envelope resumes approximately the same phase
evolution as the 328~MHz profile after the jump, until it is disturbed
by the extra component (longitude $\sim 23$\degr).  Clearly, the
situation is not as simple as the superposition of two rotating
subbeam systems offset purely in magnetic azimuth, as earlier
suggested.

\subsection{PSR B0809+74}
\begin{figure*}
\begin{center}
\begin{tabular}{cc}
\resizebox{!}{19.0cm}{\includegraphics{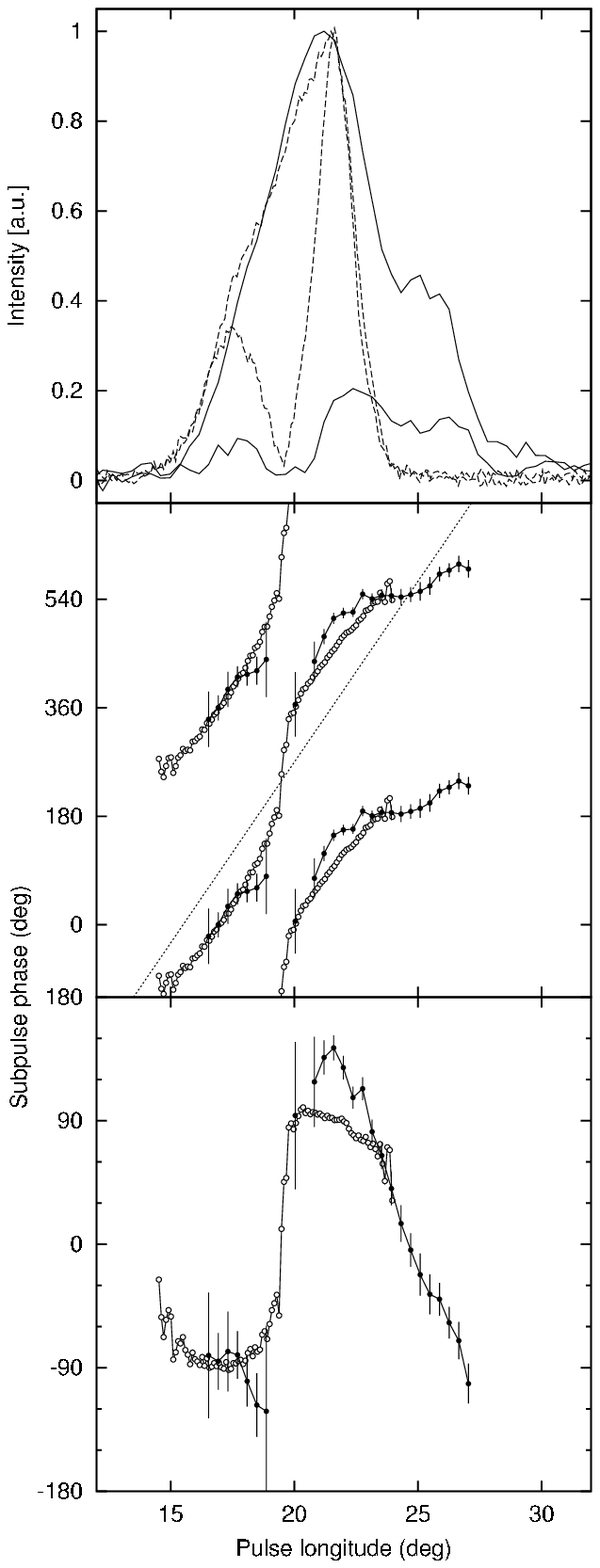}}
&
\resizebox{!}{19.0cm}{\includegraphics{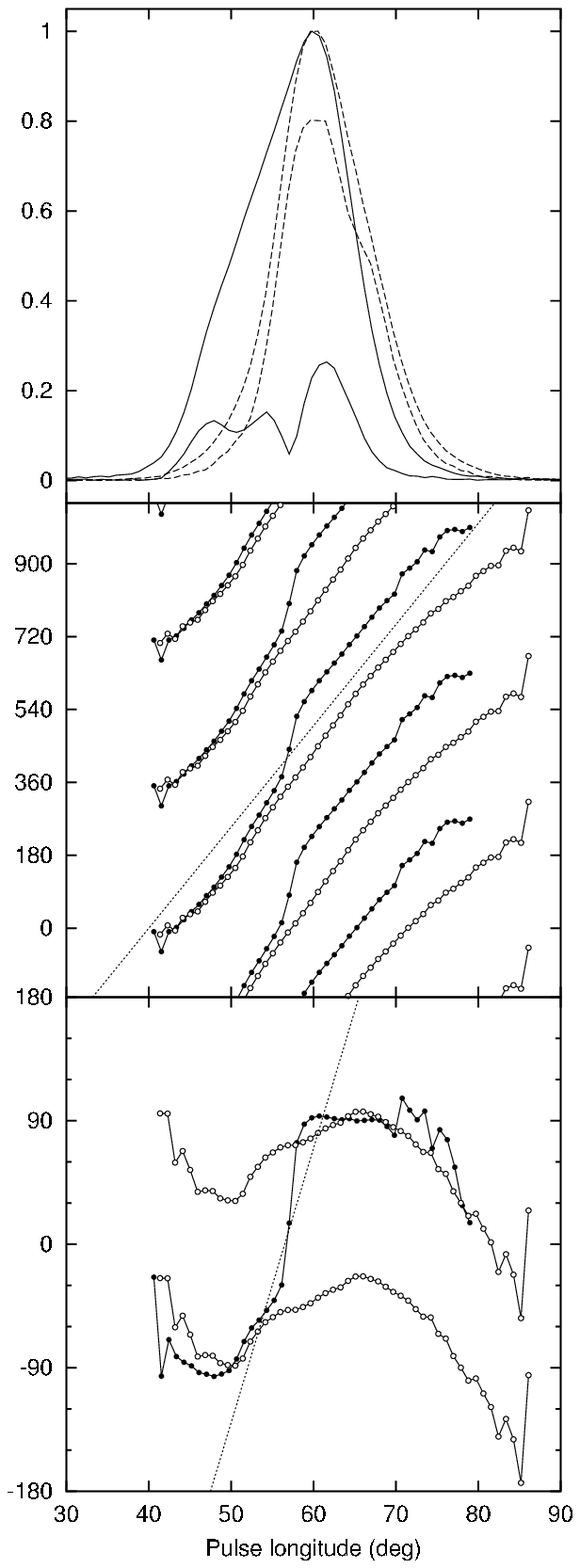}}
\end{tabular}
\end{center}
\caption{Average profiles and longitude envelopes for PSR B0320+39
(left) and PSR B0809+74 (right) at 328 MHz (dashed lines / white
circles) and 1380 MHz (solid lines / black circles).  The top panel
shows the average profile and subpulse amplitude envelope (the latter
usually being lower than the average profile).  The middle panel shows
the subpulse phase envelope, plotted repeatedly with a spacing of
360\degr to show the inherent ambiguity. Adjacent significant points
are joined by the shortest path.  The dotted line shows the nominal
phase slope ($60 \degr / \degr$ for PSR B0320+39, $25 \degr / \degr$
for PSR B0809+74). The difference between the phase envelopes and this
line is plotted in the bottom panel (for PSR B0809+74, twice, with an offset
of 120\degr\ to depict the rough consistency between the frequencies
after correction for an offset). The dotted line in the bottom
right panel indicates the slope corresponding to the $P_2$ measured by
\citet{bkk+81} at 1720~MHz. The error bars for PSR B0320+39 at
1380~MHz are twice the formal value. Error bars are omitted on other
points since they are in most places very small.  }
\label{fig:res}
\end{figure*}

The average profiles and subpulse amplitude and phase envelopes for
PSR B0809+74 at 328~MHz and 1380~MHz are shown in Fig. \ref{fig:res}.
Beginning with the 328~MHz result, we see that the subpulse phase
envelope is not as steep at the edges compared to other longitudes,
consistent with the expectations of the carousel model. However, the
high sensitivity of the data reveals that the phase behaviour is more
complicated than the smooth cubic-like forms allowed on basic
geometric grounds (ES).  The 1380~MHz phase envelope shows similar
variations in subpulse phase in the leading part of the profile, but
exhibits a striking jump in subpulse phase in the middle of the
profile, confirming earlier results of lower longitude resolution 
\citep{wbs81,pw86}.
The shape of phase envelope in the trailing part of the
profile also approximately matches that seen at low frequencies, with
an offset of $\sim 120\degr$ in relative phase compared to the leading
part. Both the large phase jump, and the smaller jump or steepening
around 52\degr\ longitude correspond to local minima in the subpulse
amplitude envelope, suggesting in our view a similar interpretation to
the phase offset in PSR B0320+39: the presence of superposed,
out-of-phase drift patterns with differing longitudinal amplitude
dependences. The cumulative effect of both phase jumps is seen to be
approximately $180$\degr, and could be explained as the consequence of
receiving rays from opposite sides of the magnetic axis, as invoked
for PSR B0329+39 (ESvL). However, in this case the but the presence an
additional ``image'' of the carousel pattern, offset by some amount
other than 180\degr, is required to explain the peak in subpulse
amplitude around 55$\degr$ longitude, and intermediate phase seen in
the region. We confirmed numerically that the major features of the
complex subpulse envelope can reproduced using three overlapping
Gaussian components with relative phases of 0, 35 and
180\degr. However, we note that this should only be taken as an
indication of the general suitability of this picture, and not a
solution for this specific set of relative phase values, for there is
considerable room for different relative phases given different
component shapes and/or a non-linear underlying phase function under
various potential viewing geometries.

These results are consistent with those of previous studies, although
for reasons outlined in the Introduction, we reach somewhat different
conclusions.  A major result of \citet{bkk+81} was that the overall
drift rate differed by a factor $\sim1.8$ between 102 and 1720~MHz, a
fact which they used to argue that the entire profile should be scaled
in longitude by this factor to correct for RFM. Performing this
scaling results in a significantly wider profile for 1720~MHz than
102~MHz, on which basis it was argued that some emission was
``missing'' at 102~MHz.  For reasons noted by \citet{gk96b} and
reiterated in the Introduction to this work, RFM under the carousel
model is not expected to alter the observed $P_2$ value the by same
factor as the profile width, so the scaling performed by
\citet{bkk+81} is not valid. Indeed, the value of $P_2$ is expected to
vary relatively little, making the observed factor of 1.8 somewhat
unexpected. However, the phase jump in the middle of the profile at
1720~MHz provides the obvious explanation for this fact.
Specifically, most pairs of subpulses in a given pulse will lie on
opposite sides of the phase jump and will thus appear closer together
than expected without the phase jump. Any method of measurement
relying on the longitudinal separation of subpulse peaks, including
the autocorrelation method of \citet{bkk+81}, will be subject to this
distortion. The dotted line in the bottom panel of \ref{fig:res}
depicts the drift rate inferred by \citet{bkk+81}, which can clearly
be seen to represent a kind of interpolation between the phases at the
peaks of the offset components we find at 1380~MHz. The drift rate
measured at 102~MHz would correspond to a horizontal line in this
plot, confirming that there is no major evolution in the drift rate
between 102 and 328~MHz.  We therefore argue that the factor of 1.8
reported in subpulse spacing between 102 and 1720~MHz is almost
entirely due to the effect of the phase offset and amplitude
windowing.

Some apparently anomalous results of \citet{dls+84} also receive
explanation from factors discussed in this work.  \citet{dls+84} state
that their observation that the drift rate  at 408~MHz is 1.5 times
faster near the profile edge edge is unexpected, and may be caused by
a non-dipole magnetic field. In fact, under the carousel model faster
drifting {\em is} expected at the profile edges (ES; see also Fig.~1
this work). We suspect that the confusion arose due to the reciprocal
relationship between ``drift rate'' (i.e. longitude displacement per
time interval) and the longitudinal rate of change of subpulse phase
(i.e. phase or time displacement per longitude interval). We also note
that the drift speed measured by \citet{dls+84} at 408~MHz actually
increases monotonically with longitude, the asymmetry of which is not
consistent with the model, with or without the noted misconception. In
any case, we have shown here that the phase envelope is considerably
more complicated than this, and yet the reduced slopes at the profile
edges indicate potential consistency with the carousel model if
offset multiple imaging is allowed. The monotonic trend seen by
\citet{dls+84} probably has more to do with the perturbations between
subpulse peaks and true points of zero phase imposed by the overall
amplitude windowing than with the underlying phase variations.

Another result of \citet{dls+84} is that the subpulses at 1412~MHz
apparently arrive 10--18~ms earlier than those at lower frequencies
(after compensating for interstellar dispersion).  Comparison of their
Figures 2 and 4 with Fig. \ref{fig:res} confirms that the offset
subpulses they recorded appeared where our proposed alignment
indicates a 120$\degr$ phase offset. This corresponds to a vertical
shift of $\sim 3.7$ periods in Fig. 4 of \citet{dls+84}, consistent
with their measurement between $1412$ and $406$~MHz, made with
absolute time alignment.  We may therefore conclude that our phase
alignment is correct and that high-frequency subpulses in the leading
part of the profile are not offset from those at low frequencies.

\subsection{Implications for Models}
The subpulse phase envelopes measured for PSR B0320+39 and PSR B0809+74
are clearly not consistent with the smooth, symmetrical curves
predicted by simple rotating subbeam models. Moreover, they
are seen to evolve strongly with frequency, and to exhibit
sharp ``jumps'' in subpulse phase. 

Further evidence for departure from the predictions of a simple
carousel model also exists in the literature. \citet{hw87} found
support for the carousel model in their detection of systematic subpulse
drift in three pulsars with ``triple'' profile morphologies. However,
the subpulse phase envelopes presented for PSR B1918+19 clearly show a
reversal in the sense of subpulse drift in the trailing
component. Under the carousel model, points of sense reversal
correspond to pulse longitudes where the line of sight is at its
furthest excursions in the magnetic azimuth coordinate.  For this to
occur within the emission beam requires the angle between the spin
axis and the observer to be comparable to the opening angle of the
beam, which would result in a very broad pulse profile ($\sim$
180\degr). This is not the case for PSR B1918+19. Indeed, given the
narrow width of pulsar beams, unless there is a trend toward alignment
of spin axes and magnetic axes, such configurations should be very
rare in the observable population, and sense reversals in general can
be taken as evidence against the presence of a simple rotating subbeam
system.

The clear cases of step-like features accompanied by reduced
modulation amplitude, seen in PSR B0320+39 and in
PSR B0809+74 at high frequencies, are suggestive of the presence of
superposed subbeam patterns that rotate together (e.g. as ``images''
of the same spark system) but are offset in the azimuthal coordinate
(ESvL). We believe that a similar effect can explain the less ordered
drift behaviour seen at other frequencies in these pulsars, and most
likely that of PSR B1918+19 also. For this to be the case, the phase
envelopes of the component images cannot be related by a constant
phase offset, requiring the corresponding subbeam patterns to have
offset centroids, or different (non-circular) shapes. To the extent
that the subbeam motions deviate from concentric circular motion in
this way, the use of subpulse patterns to determine geometric
parameters (\citealt{dr99}; ES) or to ``map'' the polar cap excitation
pattern \citep{dr99} is not possible until the transformations
effected in the images are determined, or the phase envelope is
shown to be consistent with a single rotating pattern. 

An obvious mechanism to which to appeal is the transformation to the
observer's frame, which includes aberration and retardation. The
approximate effect of these is to differentially shift the beam
patterns in the pulse longitude coordinate. However, as noted by ESvL,
this requires that the two images be emitted at points separated by
$>4500$~km, in conflict with the current consensus concerning emission
altitudes. Likewise, an altitude difference of $>2500$~km is needed to
explain the 120\degr\ phase offset seen in here PSR B0809+74 using the
combined effects of retardation and aberration.

Another possible explanation of the deviations observed is that the
magnetic field is not dipolar. \citet{dls+84} proposed such a
condition to explain the frequency evolution of the pulse profile of
PSR B0809+74, and the offset nature of subpulses at high frequencies.
Clearly this simple picture cannot now suffice, since a combination of
offset and not offset subpulses appears to be present. Perhaps, by
attributing the multiple images to multiple discrete emission heights,
some progress might be made with this model.

Allowing for refraction eases the problems by opening the possibility
of a secondary beam due to rays that cross the magnetic axis (ESvL).
However, to produce offsets other than 180$\degr$, symmetry-breaking
effects are needed. In this sense, three-dimensional refraction
calculations with full treatment of the effects of rotation on the
plasma density and flux distribution, and retardation and aberration,
should be seen as an essential next step in the understanding of
pulsar magnetospheres.

\acknowledgements {\changedd We thank the staff of WSRT for their
efforts in obtaining measurements at 272~MHz.  The WSRT is operated by
ASTRON with financial support from the Netherlands Organisation for
Scientific Research (NWO). RTE is supported by a NOVA fellowship.}

\bibliographystyle{aa}

\begin{thebibliography}{17}
\expandafter\ifx\csname natexlab\endcsname\relax\def\natexlab#1{#1}\fi

\bibitem[{Bartel(1981)}]{bar81}
Bartel, N. 1981, A\&A, 97, 384

\bibitem[{Bartel {et~al.}(1981)Bartel, Kardashev, Kuzmin, Popov, Sieber,
  Smirnova, Soglasnov, \& Wielebinski}]{bkk+81}
Bartel, N., Kardashev, N.~S., Kuzmin, A. D.~Nikolaev, N.~Y., {et~al.} 1981,
  A\&A, 93, 85

\bibitem[{Cordes(1978)}]{cor78}
Cordes, J.~M. 1978, ApJ, 222, 1006

\bibitem[{Davies {et~al.}(1984)Davies, Lyne, Smith, Izvekova, Kuzmin, \&
  Shitov}]{dls+84}
Davies, J.~G., Lyne, A.~G., Smith, F.~G., {et~al.} 1984, MNRAS, 211, 57

\bibitem[{{Deshpande} \& {Rankin}(1999)}]{dr99}
{Deshpande}, A.~A. \& {Rankin}, J.~M. 1999, ApJ, 524, 1008

\bibitem[{{Edwards} \& {Stappers}(2002)}]{es02}
{Edwards}, R.~T. \& {Stappers}, B.~W. 2002, A\&A, 393, 733

\bibitem[{{Edwards} {et~al.}(2003){Edwards}, {Stappers}, \& {van
  Leeuwen}}]{esv03}
{Edwards}, R.~T., {Stappers}, B.~W., \& {van Leeuwen}, A.~G.~J. 2003, A\&A,
  402, 321

\bibitem[{{Gil} {et~al.}(2002){Gil}, {Gupta}, {Gothoskar}, \& {Kijak}}]{gggk02}
{Gil}, J., {Gupta}, Y., {Gothoskar}, P.~B., \& {Kijak}, J. 2002, ApJ, 565, 500

\bibitem[{{Gil} \& {Krawczyk}(1996)}]{gk96b}
{Gil}, J. \& {Krawczyk}, A. 1996, MNRAS, 280, 143

\bibitem[{Hankins \& Wolszczan(1987)}]{hw87}
Hankins, T.~H. \& Wolszczan, A. 1987, ApJ, 318, 410

\bibitem[{Izvekova {et~al.}(1993)Izvekova, Kuz'min, Lyne, Shitov, \&
  Graham~Smith}]{ikl+93}
Izvekova, V.~A., Kuz'min, A.~D., Lyne, A.~G., Shitov, Y.~P., \& Graham~Smith,
  F. 1993, MNRAS, 261, 865

\bibitem[{Komesaroff(1970)}]{kom70}
Komesaroff, M.~M. 1970, Nature, 225, 612

\bibitem[{{Pr\'{o}szy\'{n}ski} \& {Wolszczan}(1986)}]{pw86}
{Pr\'{o}szy\'{n}ski}, M. \& {Wolszczan}, A. 1986, ApJ, 307, 540

\bibitem[{Radhakrishnan \& Cooke(1969)}]{rc69a}
Radhakrishnan, V. \& Cooke, D.~J. 1969, Astrophys. Lett., 3, 225

\bibitem[{Ruderman \& Sutherland(1975)}]{rs75}
Ruderman, M.~A. \& Sutherland, P.~G. 1975, ApJ, 196, 51

\bibitem[{{Vo{\^ u}te} {et~al.}(2002){Vo{\^ u}te}, {Kouwenhoven}, {van Haren},
  {Langerak}, {Stappers}, {Driesens}, {Ramachandran}, \& {Beijaard}}]{vkv02}
{Vo{\^ u}te}, J.~L.~L., {Kouwenhoven}, M.~L.~A., {van Haren}, P.~C., {et~al.}
  2002, A\&A, 385, 733

\bibitem[{Wolszczan {et~al.}(1981)Wolszczan, Bartel, \& Sieber}]{wbs81}
Wolszczan, A., Bartel, N., \& Sieber, W. 1981, A\&A, 100, 91

\end{thebibliography}

\end{document}